# RECENT RHIC IN-SITU COATING TECHNOLOGY DEVELOPMENTS*

A. Hershcovitch[#], M. Blaskiewicz, J.M. Brennan, A. Chawla, W. Fischer, C-J Liaw, W. Meng,
and R. Todd, Brookhaven National Laboratory, Upton, New York 11973, U.S.A
A. Custer, M. Erickson, N. Jamshidi, P. Kobrin, R. Laping, H. J. Poole
PVI, Oxnard, California 93031, USA
J. M. Jimenez, H. Neupert, M. Taborelli, and C. Yin-Vallgren
CERN, CH-1211 Genève 23, Switzerland
N. Sochugov, High Current Electronics Institute, Tomsk, 634055 Russia

*Abstract*

To rectify the problems of electron clouds observed in RHIC and unacceptable ohmic heating for superconducting magnets that can limit future machine upgrades, we started developing a robotic plasma deposition technique for *in-situ* coating of the RHIC 316LN stainless steel cold bore tubes based on staged magnetrons mounted on a mobile mole for deposition of Cu followed by amorphous carbon (a-C) coating. The Cu coating reduces wall resistivity, while a-C has low SEY that suppresses electron cloud formation. Recent RF resistivity computations indicate that 10 μm of Cu coating thickness is needed. But, Cu coatings thicker than 2 μm can have grain structures that might have lower SEY like gold black. A 15-cm Cu cathode magnetron was designed and fabricated, after which, 30 cm long samples of RHIC cold bore tubes were coated with various OFHC copper thicknesses; room temperature RF resistivity measured. Rectangular stainless steel and SS discs were Cu coated. SEY of rectangular samples were measured at room; and, SEY of a disc sample was measured at cryogenic temperatures.

## INTRODUCTION

Electron clouds, which have been observed in many accelerators, including the Relativistic Heavy Ion Collider at the Brookhaven National Laboratory [1-3], can act to limit machine performance through dynamical beam instabilities and/or associated vacuum pressure degradation. Formation of electron clouds is a result of electrons bouncing back and forth between surfaces, with acceleration through the beam, which can cause emission of secondary electrons resulting in electron multipacting. One method to mitigate these effects would be to provide a low secondary electron yield surface within the accelerator vacuum chamber.

At the same time, high wall resistivity in accelerators can result in unacceptable levels of ohmic heating or to resistive wall induced beam instabilities [4]. This is a concern for the RHIC machine, as its vacuum chamber in the cold arcs is made from relatively high resistivity 316LN stainless steel. This effect can be greatly reduced by coating the accelerator vacuum chamber with oxygen-free high conductivity copper (OFHC), which has conductivity that is three orders [5,6] of magnitude larger than 316LN stainless steel at 4 K. And, walls coated with titanium nitride (TiN) or amorphous carbon (a-C) have shown to have a small secondary electron yields (SEY)[7,8]. But, recent results [9] strongly suggest that a-C has lower SEY than TiN in coated accelerator tubing. Applying such coatings to an already constructed machine like RHIC without dismantling it is rather challenging due to the small diameter bore with access points that are about 500 meters apart. Although R&D has yielded some results, it is still work in progress.

___________________________________________
*Work supported by Work supported under Contract No. DE-AC02-98CH1-886 with the US Department of Energy.
#hershcovitch@bnl.gov

## DEPOSITION PROCESSES AND OPTIONS

Coating methods can be divided into two major categories: chemical vapor deposition (CVD) and physical vapor deposition (PVD). Reference [11] contains a comprehensive description of the various deposition processes; unless otherwise noted, information contained in the next two sections is referenced in [11].

Due to the nature of the RHIC configuration, only PVD is viable for in-situ coating of the RHIC vacuum pipes. First, the temperature under which coating can be made cannot be high (400$^o$C is required for some conventional CVD), since the RHIC vacuum tubes are in contact with superconducting magnets, which would be damaged at these temperatures. A second very severe constraint is the long distance between access points. Introduction of vapor from access points that are 500 meters apart into tubes with 7.1 centimeters ID would likely not propagate far and result in extremely non-uniform coating.

But these constraints also severely restrict PVD options. Obviously evaporation techniques (ovens, e-beams) cannot be used in 7.1 centimeters ID, 500-meter long tubes for the same reasons. Therefore, evaporation must be accomplished locally. One option is a plasma device on a mole that generates and deposits the vapor locally.

Presently, there are a variety of PVD methods used to deposit coatings on various substrates [11]. By definition, physical vapor deposition entails purely physical processes of evaporating materials. The vapor then condenses on the desired substrate. There is a wide variety of vapor generation techniques ranging from high temperature evaporation to sputter bombardment by electron beams, ion beams and plasma. The latter involves a discharge like RF, glow, or an arc. The long distance between access points and the need to have a mole like deposition device precludes the use of RF plasmas.

## MAGNETRON DEPOSITION STATE-OF-THE-ART

Of the plasma deposition devices like magnetrons, diodes, triodes, cathodic arcs, etc., magnetrons are the most commonly used plasma deposition devices. In magnetrons, magnetic fields are utilized to confine electrons that generate high density plasma (usually argon or xenon) near the surface of the material that is being sputtered. Major advantages of magnetron sputtering sources are that they are versatile, long-lived, high-rate, large-area, low-temperature vaporization sources that operate at relatively low gas pressure and offer reasonably high sputtering rates as compared to most other sputtering sources. Because of these superior characteristics magnetron sputtering is the most widely used PVD coating technique. Although arc discharges operate with higher intensity, they require the use of special filters [12] to eliminate macroparticles that reduce the net deposition rate to those of magnetrons.

Typical coating rates by magnetrons (w/argon gas) are 5 Å/sec for a power of 10 W/cm$^2$ on the magnetron cathode, though with intense cooling cathode power of 20 W/cm$^2$ is achievable.

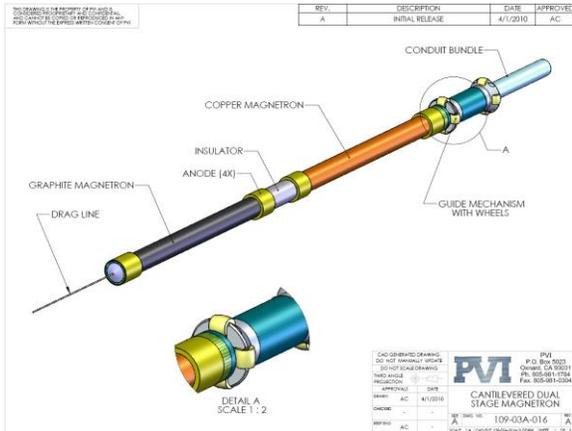

Figure 1: Diagram of the deposition device based on dual stage magnetrons.

## PLANNED DEPOSITION TECHNIQUE

The ultimate objective is to develop a plasma deposition device for *in-situ* coating of long, small diameter tubes with about 5 - 10 μm of Cu following by a coating of about 0.1 μm of a-C. Figure 1 is the original scheme of a plasma deposition technique based on staged magnetrons. Plasma deposition sections consist of two, connected through an insulator, cylindrical magnetron devices. The first magnetron stage has oxygen free high conductivity copper cathode, while the second stage has a graphite cathode. Internal ring permanent magnets form the magnetic field. Magnetron assembly is to be mounted on a carriage (*mole*), which is to be pulled by a cable assembly driven by an external motor. To accommodate for any diameter variances, including bellow crossing, the carriage will have a spring-loaded guide wheel assembly. Spool drive mechanism is shown in figure 2. A dragline, which is attached to end (opposite to the carriage) of the graphite cathode, is used to initially pull the magnetron assembly and cable bundle to the end, where coating begins. The dragline, which is also motor driven, is a strong thin cable made of either high-tensile fishing line, or Teflon sleeved (Teflon coated) Inconel or equivalent. Should there be evidence that either the Teflon or the fishing line live any residue, a pure metal line is to be used. During coating, the magnetron assembly and cable bundle are pulling the dragline (in a direction opposite to, which the dragline pulled on the magnetron assembly and cable bundle).

If needed, a brushless DC servo-motor driving 4 rows of internal wheels moves the carriage, which has position feedback, assists carriage motion. Cable for pulling mole identified: ~ 6 mm diameter stranded SS with a Teflon sheath. This type of cable is typically used in aircraft for flexible linkage with the various airfoil surfaces (rudder, flaps...etc.). It is very strong (20K tensile) with low elongation.

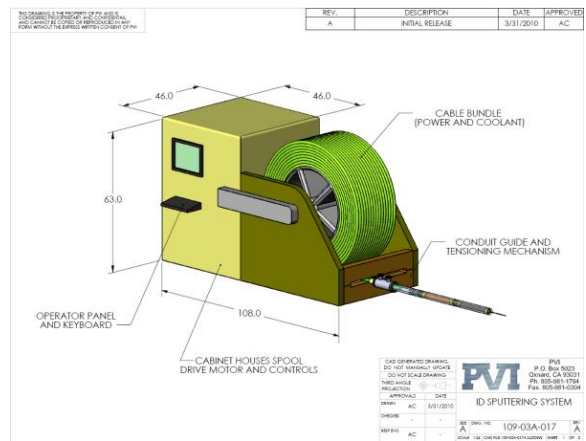

Figure 2: Perspective view of spool drive mechanism.

Based on the fact that magnetrons with 2.1 meter long cylindrical cathodes exist in commercial systems [13], in a previous paper[14], it was assumed the copper magnetron section can be 2 m long. And at a Cu coating rate of 5 Å/sec (though much higher rates were achieved), it would take 2.78 hours to deposit 5 μm of Cu, i.e., close to 3 hours to move one cathode length. With a 2 meter long cathode it would take 695 hours (or 29 days; a fraction of a typical RHIC shutdown period) to coat 500 m. And 2 m Cu cathode would not need reloading.

But magnetron weight would limit single deposition device length to about 50 cm. Consequently, the technique is to involve one of two options: multiple magnetrons in a train like assembly, having a total exposed cathode length of 2+ meters, as shown in figure 3, or magnetrons with reloading provisions, which would require access bellows.

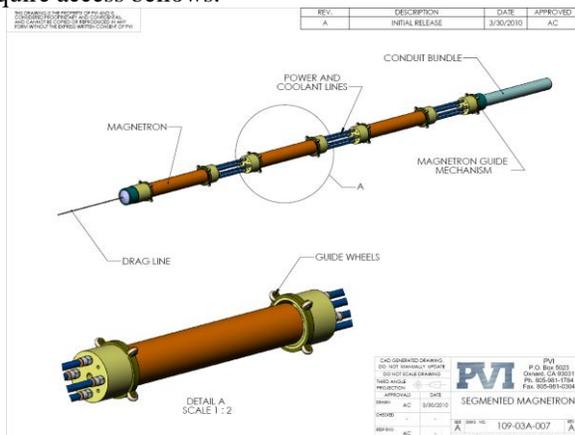

Figure 3: Sketch of multiple magnetrons.

Some of RHIC bellows can be replaced with access bellows, as shown in figure 4, to enable cathode reloading.

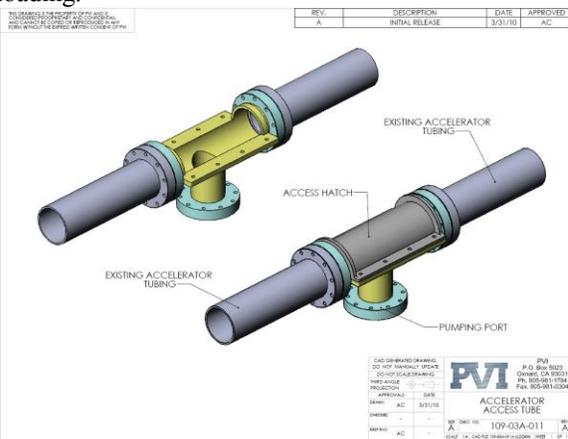

Figure 4 Drawing of access bellows.

If support wheels can be utilized, multiple magnetrons in a train like assembly would work. Presently replacement cathodes utilizing access bellows is not the leading option. Taking few magnets out and coating a series of magnets at a time is being considered. No final decision has been made.

## MAGNETRON OPERATION

A mobile magnetron, shown in figure 5, with a 15 cm long cathode was designed, fabricated, and tested to coat 32 cm long samples of RHIC cold bore tubes with up to 6.1 μm with OFHC at an average coating rate of 30 Å/sec. Copper deposition rates were measured with a 6 MHz crystal rate monitor. A coated sample is shown in Fig. 6.

Experiments were performed in a deposition chamber (shown in figure 7), in which 30 cm long RHIC cold bore samples were mounted. Initially, there were discharge ignition difficulties (operating on the LHS of the Paschen curve) with the magnetron inside a pipe in relatively big box. Discharge intensity and coating rates were dominated by edge effects. Additionally, there was very poor copper utilization due to very uneven longitudinal discharge intensity, as it can be seen in figure 8 (on the right), which resulted in narrow waists that compromised magnetron integrity (due to magnetic field shape and magnet variation; 2 were 80+ mT others 50-60 mT). But, plasma discharge and deposition are azimuthally uniform (the left of figure 8). Nevertheless edge effects would not be an issue in a long tube.

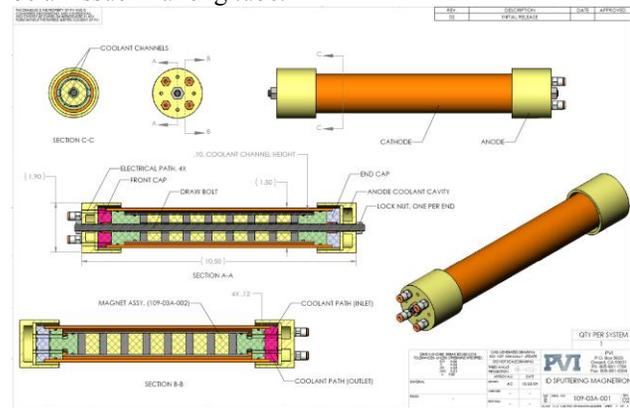

Figure 5: Complete drawing of the experimental magnetron.

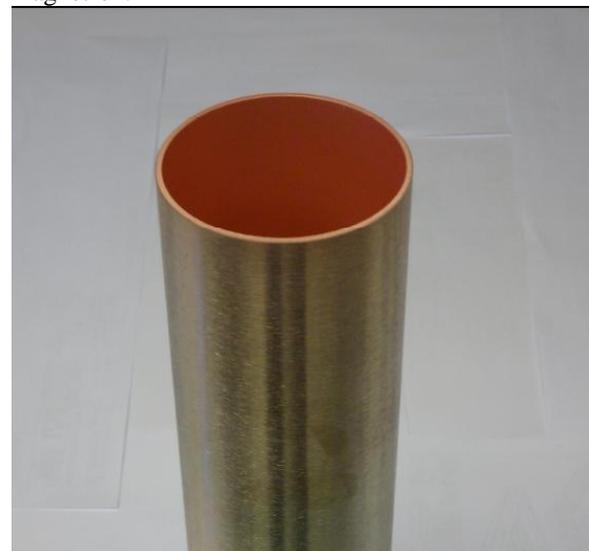

Figure 6: Copper coated RHIC tube sample.

First coated samples were 30 cm long samples of RHIC cold bore tubes with various thicknesses in the range of 2.5 μm - 6.1 μm.

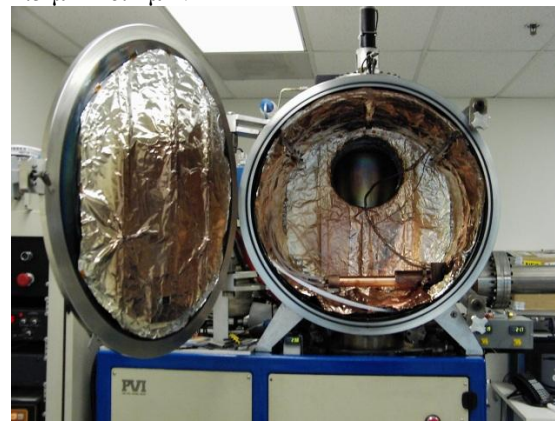

Figure 7: Deposition chamber; magnetron and RHIC tube sample are on the bottom.

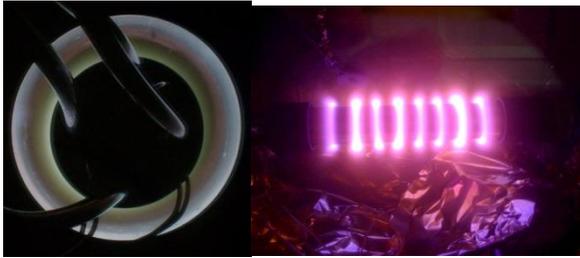

Figure 8: Photo of argon plasma between magnetron surface and tube (left); power, cooling, and instrumentation feed visible; plasma discharge and deposition are azimuthally uniform. But axially discharge is non-uniform (right).

## COATING ADHESION PROBLEMS

First coatings with DC power had poor adhesion. Occasionally coating with good adhesion was achieved with AC at 40 kHz (square wave) deposition. But, it was inconsistent, adhesion did not always meet rigorous industrial standard (tape; nail).

Pre-coating was tried to enhance adhesion. Nickel (top industrial choice) is magnetic and therefore could not be used. Chrome, which is hard to sputter, causing very uneven erosion and poor copper cathode utilization, was not an option either. So titanium pre-coating was tried. A bi-metal magnetron, (figure 9) was fabricated.

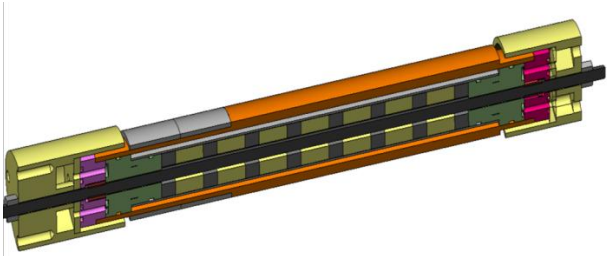

Figure 9: Drawing of bi-metal titanium-copper magnetron.

After considerable effort, successful operation was accomplished. Although copper to titanium adhesion was excellent, titanium adhesion to stainless steel was poor. Nevertheless, useful experience was gained in case simultaneous copper and carbon operation are needed.

## SOLUTION: DISCHARGE CLEANING

The adhesion problem was solved with discharge cleaning. The first step is to apply a positive voltage (of about 1 kV) to the magnetron or a separate cleaning anode and to move the discharge down the tube at a pressure of over 2 Torr. So far it worked well with the existing magnetron (for long tube cleaning there is concern of discharge cleaning debris affecting the copper cathode).

The second step is the conventional deposition step at a pressure of about 5 mTorr. Initial good adhesion was accomplished with Ti pre-coating; later with direct copper coating. No need for pre-coating!

## REQUIRED COATING THICKNESS

The needed copper coating thickness is determined by RF resistivity requirements. Computations indicated[15] that the combined effects of low temperature and large magnetic fields will yield a net reduction in room temperature resistivity of RRR=50 in the copper coating. The mean free path of conduction electrons is 2 μm, which is equal to the skin depth at 20 MHz. It is therefore prudent to include the anomalous skin effect when calculating the effect of the coating. When this is done it is found that 10 μm of copper should be acceptable for even the most extreme future scenarios.

Studies that were made for thick copper coatings[16-18] of a few micrometers or more have shown that the upper layers of the coatings have columnar and other grain structure rather than crystalline. Thus, those layers might have a low SEY like gold black. Therefore, SEY of thick copper coatings need to be measured, since a low SEY may eliminate the need for a-C coating.

## COATING GENRES

Theoretically, the coating structure should depend on magnetron discharge conditions. Hypothetically, therefore, copper coatings with crystalline or columnar and other grain structures can be deposited with the proper choice of magnetron discharge parameters. Furthermore, different layers having different structures can be deposited successively. In principle therefore, a thick layer of nice crystalline like structure can be deposited on the RHIC cold bore tube to lower RF resistivity, on top of which, a thin copper layer with columnar and other grain structure is deposited to lower SEY.

Visually, deposited copper with crystalline like, high density, structure is supposed to be shiny, while deposited copper with columnar and other grain structure should be matte in appearance, like gold black. In principle copper coating, which is matte visually, should have low SEY.

Four magnetron operating modes work well in the figure 7 geometry (with the magnetron inside the tube): low pressure (5 mTorr or lower), high pressure (20 to 40 mTorr), AC or DC power. But only deposition at high pressure with AC power can result (but not always does) in visually matte copper coating of stainless steel samples. Shiny and matte coated samples are shown in figure 10.

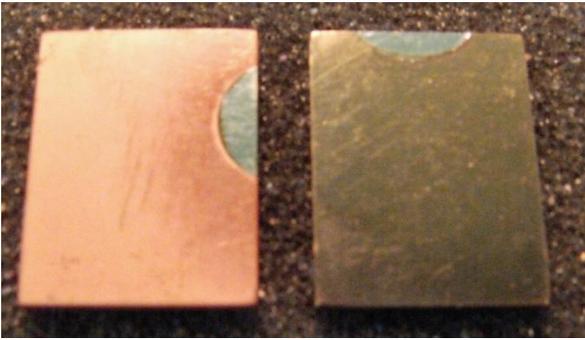

Figure 10: Rectangular copper coated stainless steel samples that are visually shiny (left) or matte (right).

## RF RESISTIVITY MEASUREMENTS

The first coated samples were those of three 32 cm long RHIC stainless steel tubes. Without discharge cleaning, tubes were coated with 2.5 μm to 6.1 μm OFHC at 20 mTorr with AC power. Coating was axially non-uniform (thicker at edges) and matte in appearance suggesting higher resistivity and lower SEY.

Room temperature RF resistivity of one of the coated samples (shown in figure 6) 2.5 μm or about 4.5 to 5 μm (coating thickness marking was lost), was close to copper at 180 MHz with coating that's far from ideal.

To rectify the problem of non-uniform RHIC stainless steel tube coating due to edge effects, 49 cm long tubes were coated, out of which the center 32 cm were cut out for additional RF resistivity testing. Three tubes with OFHC coatings, with thicknesses of 2 μm, 5 μm, and 10 μm, were made.

Additional measurements were made using resonant cavities. For a fixed geometry the quality factor of a resonant cavity is proportional to the inverse of the real part of the surface resistivity [15]. To test the coatings we measured the quality factor of a resonant cavity made of solid copper and the quality factors with coatings of 2, 5, and 10 μm of copper on a stainless steel substrate. The ratio of the quality factors should equal the inverse ratio of the surface resistivities. For reference the surface impedance of a layer of thickness $\tau$ and conductivity $\sigma_1$ on a substrate of conductivity $\sigma_2$ is

$$\frac{E}{H} = (1-i)\omega\mu_0\delta_1\frac{1+X}{1-X}$$

Where

$$X = \frac{\delta_2 - \delta_1}{\delta_2 + \delta_1}\exp\left[-2(1-i)\frac{\tau}{\delta_1}\right]$$

and $\delta = \sqrt{2/\omega\sigma\mu_0}$ is the skin depth for a given material and frequency. Figures 11, 12 and 13 show measured Q ratios as a function of frequency as well as the theoretical values assuming that the thickness of the coating was correct but that its conductivity could be different from that of pure copper for Cu coating thicknesses of 10, 5, and 2 μm respectively.

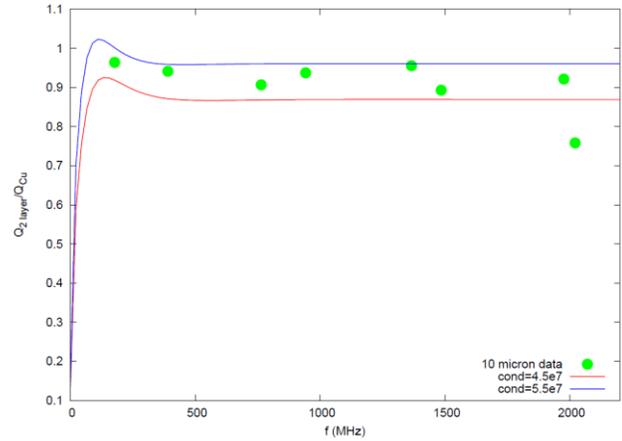

Figure 11: Ratio of SS tube coated with 10 μm of copper to pure copper tube versus frequency; experimental data is represented by green dots; red and blue lines are theoretical values based on σ of 4.5 and 5.5 x 10$^7$ mho/meter respectively.

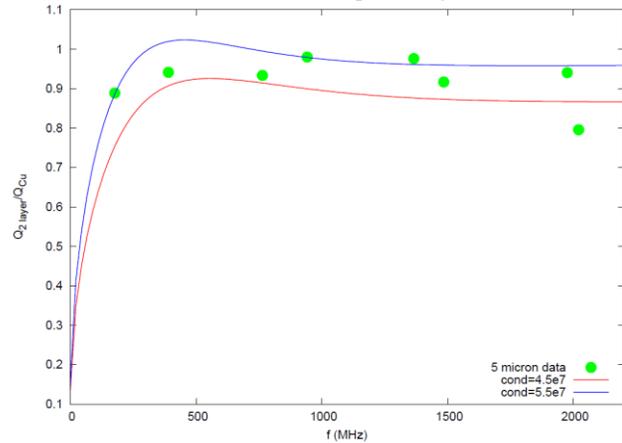

Figure 12: Ratio of SS tube coated with 5 μm of copper to pure copper tube versus frequency; experimental data is represented by green dots; red and blue lines are theoretical values based on σ of 4.5 and 5.5 x 10$^7$ mho/meter respectively.

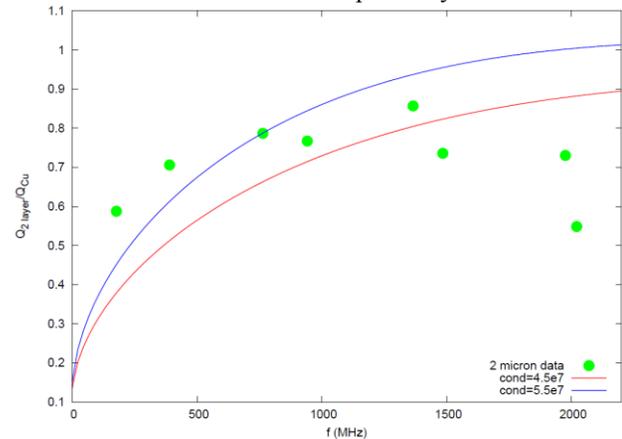

Figure 13: Ratio of SS tube coated with 2 μm of copper to pure copper tube versus frequency; experimental data is represented by green dots; red and blue lines are theoretical values based on σ of 4.5 and 5.5 x 10$^7$ mho/meter respectively.

As it can be seen from figures 11 – 13, the best value for the conductivity of the surface layer is between 4.5 and 5.5 x $10^7$ mho/meter. Pure copper has a value of 5.96 x $10^7$ mho/meter. Thus, based on these measurements the conductivity of the copper coating is between 75.5% and 92.3%, or about 84% of pure copper.

Nevertheless resistivity at cryogenic temperature will most likely be different (it must be measure in a system that's not yet available).

## COATED SAMPLE PREPARATION

For SEY measurements three 29 mm diameter stainless steel discs and three 15x20 mm rectangular samples were copper coated with thicknesses of 2 μm, 5 μm, and 10 μm. SEY of the rectangular samples were measured at room temperature; while SEY of one disc was measured at cryogenic temperatures; SEY measurement were performed at CERN.

Given that coating characteristics vary with deposition parameters, an effort was made to ensure that coated samples are made under magnetron operating parameters that are as close as possible to operation inside the RHIC cold bore tubing. To that end, coating of disc and rectangular samples were made with a magnetron inside a RHIC cold bore tubing section with three holes, as shown in figure 14 below.

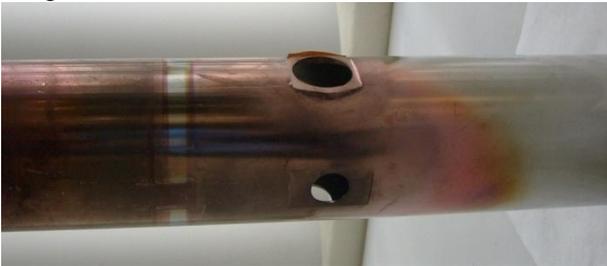

Figure 14: Tubing, in which samples for SEY measurements are prepared.

The crystal rate monitor is mounted in one hole to measure deposition rates and deposition thickness. Two identical samples are mounted in the other apertures for deposition. The magnetron is then inserted and operated inside the tube to coat samples under conditions that are as close as possible to magnetron copper coating inside the RHIC cold bore tubing. After coating is completed, adhesion tests are performed on one sample, while the other sample is sent out for SEY measurements.

## ROOM TEMPERATURE SEY MEASUREMENTS

All SEY measurements were performed at CERN on OFHC copper coated rectangular stainless steel samples, with Cu coating thicknesses of 2 μm, 5 μm, and 10 μm. First measured samples were prepared with DC power at low pressure, and were shiny in appearance. Measurements were performed without any cleaning or baking. Maximum room temperature SEY ($\delta_{max}$) of 1.65 at an energy of 332 eV for the 10 μm coated sample, and $\delta_{max}$ increased as the coating thickness decreased ($\delta_{max}$ = 1.78 for 2 μm coating) as expected, since thicker coatings are more likely to have columnar and other grain structure.

Encouraged by those results, additional samples were prepared with DC power at low pressure (5 mTorr), except for the upper 0.3 μm layer, which was deposited at high pressure (35 mTorr) with AC power. The resultant SEY measurements, however, were totally unexpected: instead of lower SEY, higher values for $\delta_{max}$ were recorded. Furthermore, $\delta_{max}$ increased for increasing (rather than decreasing) coating thickness. At room temperature $\delta_{max}$ = 1.79 for 2 μm thick coating at an energy of 382 eV, which increased to $\delta_{max}$ = 1.86 for 10 μm thick coating also at an energy of 382 eV.

## CRYOGENIC TEMPERATURE SEY MEASUREMENTS

Driven by the logic that the bulk of the copper coating should be crystalline like, high density coating for good conductivity followed by a relatively thin layer of columnar and granular visually matte copper, 2 μm, 5 μm, and 10 μm disc samples were prepared with DC power at low pressure (5 mTorr), except for the upper 0.3 μm layer, which was deposited at high pressure (35 mTorr) with AC power. Samples were sent to CERN; interesting results were obtained.

Initially, SEY of the rectangular samples were measured at room temperature on the CERN SEY system connected to their XPS, i.e. a baked vacuum system (1 x $10^{-9}$ mbar) which enables sample transfer from air to UHV; $\delta_{max}$ results were practically identical to the last room temperature SEY measurements, i.e., room temperature $\delta_{max}$ = 1.79 for 2 μm thick coating at an energy of 382 eV, which increased to $\delta_{max}$ = 1.86 for 10 μm thick coating also at an energy of 382 eV.

Next, the disc with the 2 μm thick coating was mounted in air on the CERN cryogenic head for 9 K SEY measuring device. First SEY was measured at 300 K before any bake-out; room temperature $\delta_{max}$ = 2.15 at an energy of 300 eV. The vacuum system and the disc were then baked at 150 C; room temperature $\delta_{max}$ dropped to 1.55 at energy of 250 eV. The sample disc was then cooled to 8.6 K; SEY measurements reveal $\delta_{max}$ = 1.53 at energies of 250 to 300 eV.

## SEY RESULTS DISCUSSION

At first glance, SEY results are unexpected, since the earliest shiny (crystalline like high density) samples exhibited lower SEY than the matte (in principle with columnar and granulous upper layers) samples. Furthermore, at larger coating thickness one would expect also a larger roughness and hence a lower SEY, whereas the observed difference is close to the experimental accuracy (+/- 0.03 on the SEY value).

It is well known that the SEY value is sensitive to surface contamination [19]. This might explain differences between different series (the shiny and the matte one) which were transported in different times. Along the same line, eliminating contamination can reduce copper SEY to a point where electron clouds would not form. In the present case the samples with 2 µm thick coating introduced in the unbaked system showed a higher SEY compared to those measured in the baked vacuum system. We observed in the past on carbon coatings that the exposure to an unbaked system can increase the SEY. The mechanism of this behaviour is still unclear. In addition baking the sample with the 2 µm thick coating is sufficient to reduce its SEY from 2.15 to 1.55. Also in this case we are facing a cleaning of the surface through desorption of water related and hydrocarbon species, as in a common bake-out of a vacuum system. Scrubbing will further reduce its SEY, as it is known from the literature data for copper [20].

## NEAR-TERM PLANS

A magnetron with a 50 cm long copper cathode is being designed and fabricated (cooling and weight limits the length). A Tesla coil or a beta emitter (Ni-63) is to be utilized to initiate/maintain discharge. To increase cathode lifetime, thicker cathode (x2), stronger magnets, and movable magnet package are used.

A new test stand comprising of full-size dipole vacuum tube with removable testing middle section, two types of RHIC bellows, differential pumping for magnetron insertion is being setup as shown in figure 15.

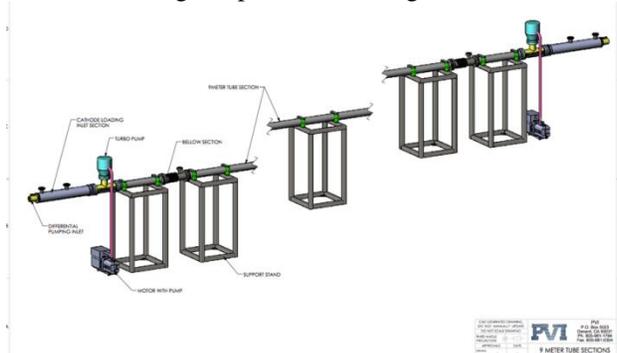

Figure 15: Diagram of new test stand.

Among the tests planned for the new test stand is discharge cleaning in confined tube: debris, oxidation, and hydrocarbons removal (Ar-$O_2$ and Ar-$H_2$ glow discharge).

Longer term plan is to perform magnet quench tests on copper coated RHIC cold bore tubing.

## DISCUSSION

From its inception, some aspects of this project and a few of its tasks seem daunting. To begin with the geometry of a RHIC in-situ coating configuration, with a target to substrate distance of 3 cm or less, is rather challenging, when compared to commercial coating equipment, where the target to substrate distance is 10's cm; 6.3 cm is the lowest experimental target to substrate distance found in the literature. Additionally, the magnetron developed here provides unique omni-directional uniform coating.

A number of challenging hurdles were anticipated[14], some of which materialized, while other unforeseen problems, like adhesion required substantial effort for solution. Eventually, a good reliable coating method with good adhesion was developed. A number of additional important milestones and achievements were reached. Cable for pulling the mole is identified; and, solutions were found for engineering issues like bellow crossing and good copper utilization. The RF resistivity of coated RHIC tube samples was found to be close to copper; nevertheless, RF resistivity measurements must be repeated at cryogenic temperatures.

Since well-scrubbed bare copper can have its SEY reduced to 1, it does not seem, at this point, prudent to further pursue copper coating with matte finish, especially since deposition at high pressure with AC power is a slower coating process. Therefore, the best approach for RHIC at this point is to coat at low pressure with DC power resulting shiny crystalline like high density coating, and to develop an in-situ plasma discharge cleaning.

These are encouraging results but, there are still more questions to be answered and challenges to overcome. But, no obstacles appear insurmountable at this point.

## ACKNOWLEDGEMENT

Notice: This manuscript has been authored by Brookhaven Science Associates, LLC under Contract No. DE-AC02-98CH1-886 with the US Department of Energy. The U.S. Government retains, and the publisher, by accepting the article for publication, acknowledges, a world-wide license to publish or reproduce the published form of this manuscript, or others to do so, for the United States Government purposes.